\documentclass[a4paper,10pt,nofootinbib]{revtex4-2}

\usepackage{caption}
\usepackage{subcaption}

\usepackage{graphicx}  % standard LaTeX graphics tool;
\usepackage{amsmath}   % For getting proper math eqns
\usepackage{amssymb}   % Used for getting various symbols eg. gtrsim, lesssim etc.
\usepackage{bm} % For bold math (esp of lower greek letters)
\usepackage{dcolumn}% Align table columns on decimal point
\usepackage{color}
\usepackage{mathrsfs}
\usepackage{amsfonts}
\usepackage{varioref}
\usepackage{mathrsfs}
\usepackage{graphicx}
\usepackage{latexsym}
\usepackage{amsmath}
\usepackage{amssymb}
\usepackage{textcomp}
\usepackage{amsbsy}
\usepackage{graphics}
\usepackage{epstopdf}
\usepackage{color}
\usepackage{enumitem}
\usepackage{array,multirow} %Tables
\usepackage{float}
\usepackage{extarrows}
\usepackage{url}
\usepackage{caption}

\input epsf

\usepackage{hyperref}
\hypersetup{
    colorlinks=true,
    citecolor=blue,
    linkcolor=blue,
    filecolor=magenta,      
    urlcolor=magenta,
}

\def\input@path{{./}}

\begin{document}

\tolerance=5000

\title{I-Love-Q relations for Neutron Stars with Dark Energy}
\author{Simone~D'Onofrio$^{1}$ \, \thanks{donofrio@ice.csic.es}} \affiliation{
$^{1)}$ Institute of Space Sciences (ICE, CSIC) C. Can Magrans s/n, 08193 Barcelona, Spain}

%\date{}

\tolerance=5000

\begin{abstract}

The influence of a dark energy fluid on the equation of state of neutron stars is investigated. A detailed analysis is conducted for such models, including the computation of the moment of inertia, the quadrupole moment, and the tidal Love number. The results demonstrate that these quantities are interconnected through the well-known equation of state independent I-Love-Q relations. This work extends the applicability of these universal relations to a broader class of neutron star models.
\end{abstract}

\maketitle

\section{Introduction}
It is currently estimated that the majority of the energy inside our universe is in the form of dark energy (DE) \cite{2020}. This energy is responsible for the accelerating expansion of the universe, and its properties are still an open problem. In the literature, it is possible to find many models that describe this effect using different approaches \cite{Li_2011,Bamba_2012,YOO_2012}. Modified gravity theories may be the sources of DE, since they give models that explain the inflationary period and the early-time cosmological acceleration (for reviews see \cite{Nojiri_2011,Nojiri_2017}).
Usually, this problem, which arises from cosmology, is studied in the same context, but some studies have been proposed to tackle it from an astrophysical point of view. Since the account of modified gravity leads to modified Tolman–Oppenheimer–Volkoff (TOV) equations, the phenomenology of neutron stars (NSs) is of major interest in the last years \cite{Astashenok:2020qds,astashenok2023compactstarsdarkenergy,Astashenok:2021xpm,Astashenok:2014nua,Astashenok_2020,Astashenok_2020_supermassive}. The different approaches can lead to interesting results and new possibilities to discern between the multiple models of DE. \\
Due to the recent developments in the study of NSs, we decided to take them as a laboratory to study the validity of different DE models and the properties that stars immersed in a DE fluid can have. The multi-messenger analysis of such objects led in the last years to important discoveries and accurate results for the NSs observables. In particular, recent results have been obtained through the study of GWs and electromagnetic surveys, such as the NICER mission \cite{Miller_2021}, which gives an accurate estimation of the radius of such objects. Multiple studies have been performed to study the effect of DE on such objects. In particular, the mass-radius relations have been deeply investigated and also the stability of such objects \cite{Carter_2005, Hess_2015}. The study of the effect of DE has been addressed also to the other class of compact objects, black holes (BHs), giving interesting results for the gravitational collapse \cite{Chan_2011} and non-singular models \cite{Dymnikova:1992ux}. \\
Among the different properties that can be studied for such modified objects, it seems that in the literature it is not deeply analyzed the validity of the universal relations of NSs with DE component. To our knowledge the only study on the topic is the study NSs with the extended Chaplygin gas equation-of-state for DE \cite{galaxies13010013,Pretel_2024}. 
The universal relations were firstly pointed out in 2013 by Yagi \& Yunes \cite{Yagi_2013,Yagi_2013_Science} (see \cite{Yagi_2017} for a review) and represent a set of relations between the moment of inertia, the quadrupole moment and the tidal Love number (TLN) of NSs.
These relations state that any of the two variables of the trio are related by an EoS independent function, giving the possibility to infer one of the quantities from another, breaking degeneracies between astrophysical parameters, and to test the validity of general relativity (GR).
This universality has then been linked to the incompressible limit of dense matter by \cite{Sham_2015}, and it is valid for the first four multipoles \cite{Pappas_2014}. 
The generalization to rapid rotating stars has shown a breakdown of these relations \cite{Doneva_2013}. However,  in this case, we can recover the universality when the rotation is a function of a suitable dimensionless parameter  \cite{Chakrabarti_2014,Stein_2014}. The formalism has been then extended to anisotropic stars \cite{Yagi_2015}, making the relations valid for a larger class of stars.  \\
These relations seem to be insensible to the EoS of the object and have been proved for a large class of models for the internal structure \cite{Haskell_2013} and exotic compact objects \cite{Berti_2024}. In particular, many studies have been performed for a large class of dark matter models \cite{Maselli_2017,Wu_2025}, providing always an agreement of such relations. Also NSs in modified gravity theories have been analyzed \cite{Pani_2014,Sham:2013cya}, showing always a good independence of such relations from the underlying models. The reason behind the universality of such relations is still not well understood. One possible explanation can be that these relations mainly affect the outer layer of the star, where the pressure, and then the energy density, of the star is small. All the models almost converge at low pressure-density values, independently from the EoS. For the reasons above, we considered interesting to investigate if the addition of a DE contribution to some of the most used EoS for NSs will generate deviations from these relations, or on the opposite, they are independent also for such a change in the model. Studying these relations for different models of DE can help in discriminating between them and to understand better the reason of such invariance.
Among all the DE models, we decided to study these relations for the standard $\Lambda$CDM model and for the quintessence model, due to their importance in the field of cosmology and to their analytical simplicity. In the future, it will be interesting to extend the same analysis to different DE EoS.
It will be interesting to study these relations for dynamical DE models, in light of the recent result of the DESI collaboration \cite{desicollaboration2025desidr2resultsii}, for example the $w_0w_a$CDM model or the entropic DE. Also the extension to the anisotropic star case could be interesting, where the DE component could in principle accelerate the deviation of the universality with respect of an increase of the amount of anistropy. If this contribution is enough relevant, it could be possible to obtain a breakdown of the relations for realistic anisotropies.
\\
\\
The paper is structured in this way. In Section \ref{sec:framework} we introduce the important quantities needed for the study of the universal relations. In Section \ref{sec:I-Love-Q} we report the results obtained from the integration of the field equations and the I-Love-Q relations. In Appendix \ref{sec:appendix} are presented all the field equations and a detailed explanation of how the numerical analysis has been performed. Finally, in Section \ref{sec:conclusions} we summarize the results obtained. \\
In the following, we used the geometrized units $G=c=1$.

\section{Framework}\label{sec:framework}
The description of rotating stars is based on the Hartle-Thorne work \cite{Hartle:1967he,Hartle:1968si}, in which they provide the most general metric for a slowly rotating stationary axisymmetric distribution of matter which has the following expression
\begin{equation}
\begin{aligned}\label{HTmetric}
    ds^2 =& - e^{\nu} \left[ 1 + 2 \epsilon (h_0 + h_2 P_2)\right] dt^2 + \frac{1 + 2 \epsilon^2 (m_0+ m_2 P_2)/(r-2m)}{1-2m/r} dr^2 \\ &+ r^2 \left[1 + 2 \epsilon^2 (v_2 - h_2)P_2\right]\left[d\theta^2 + \sin^2 \theta (d\phi - \epsilon \omega dt ) ^2\right] \ ,
\end{aligned}
\end{equation}
where $P_2 = P_2(\cos \theta) = (3\cos^2\theta-1)/2$ is the 
Legendre polynomial. The metric is written as an expansion in the angular velocity $\Omega$, of which we kept explicit the order introducing the infinitesimal parameter $\epsilon$. 
The slow rotating configurations are obtained for values of the angular velocity $\Omega$ much smaller than the Keplerian frequency of an object, so that
\begin{equation}
    \Omega \ll \sqrt{\frac{G M}{R^3}} \ ,
\end{equation}
where, just in this formula, it has been preferred to include the dependence on the gravitational constant. In this limit, $\Omega$ is small enough to have small fractional changes in pressure, energy density and gravitational field \cite{Hartle:1967he}. It has been decided to fix the angular velocity of all the stars to be $\Omega= 300 \, \text{Hz}$, which respects this condition for all the simulated stars as can be seen from Figure \ref{fig:mass_frequency}.
The slow rotating condition is well-justified for old NSs, which are known to rotate with a small frequency relative to their masses, especially considering second-order contributions. This framework breaks down for young NSs. \\
All the functions introduced are only functions of the radial coordinate due to the static hypothesis. The zero-order ones are $\nu$ and $m$ and reproduce the standard non-rotating, spherical and static metric. At first order we only have the angular velocity $\omega$, which will be written in terms of $\overline{\omega} = \Omega-\omega$, since the equations of motion will be only dependent on this variable.
The left ones are second-order functions, which are $h_0$, $h_2$, $m_0$, $m_2$ and $v_2$.\\
We will consider objects that are uniformly rotating. In this case, the matter sector will be represented by a perfect fluid. For such fluid, the stress-energy tensor is given by
\begin{equation}\label{Tmunu}
    T_{\mu\nu} = (\rho + P) u_\mu u_\nu + P g_{\mu\nu} \ ,
\end{equation}
with $\rho$ and $P$ being the zero-order energy density and pressure of the fluid. The four-velocity $u^\mu$ for a rotating object is found to be
\begin{equation}
    u^\mu = \left( u^0,0,0,\epsilon\Omega u^0\right) \ ,
\end{equation}
and has to respect the condition $u_\mu u^\mu = -1$, which to second order in the rotation gives the solution for the zero-component of the four-velocity
\begin{equation}
    u^0 = \frac{1}{\sqrt{-(g_{tt} + 2\epsilon\Omega g_{t\phi} + \epsilon^2 \Omega ^2 g_{\phi\phi})}} \ .
\end{equation}
The rotation of the star will cause a second order deformation of the star surface resulting in a change in the matter distribution and therefore of the fluid quantities. In the original papers, Hartle-Thorne \cite{Hartle:1967he,Hartle:1968si} formalized this in a change of the radial coordinate. This is equivalent to changing the fluid quantities to 
\begin{align}
    p &= P_0 + \epsilon ^2 ( \rho_0 + P_0)(p_0+p_2P_2)\\
    \rho &= \rho_0 + \epsilon ^2 ( \rho_0 + P_0)\frac{\partial \rho_0}{\partial P_0}(p_0+p_2P_2) \ .
\end{align}
With this formalism, will be described both the effects of the deformation of the star due to its rotation and the deformation caused by an external tidal field. More details on the field equations and the procedure to integrate them are given in Appendix \ref{sec:appendix}.

\subsection*{Dark Energy stars}
We will consider the contribution of a DE component to the EoS of the fluid of the star. Introducing the DE pressure and energy density
\begin{align}
    P &= P_m + P_d \\
    \rho &= \rho_m+\rho_d \ .
\end{align}
Following \cite{astashenok2023compactstarsdarkenergy}, we will use an illustrative model of coupling between standard matter and DE of the type
\begin{equation}
    \rho_d = \alpha \rho_m e^{-\frac{\rho_s}{\rho_m}} \ ,
\end{equation}
so that we have an appreciable coupling starting from a characteristic energy density cut-off $\rho_s$, chosen to be $\rho_s = 5 \times 10^{14}  \text{ g/cm}^3$. This will reproduce the known property of DE of not interacting with standard matter for small densities. The constant $\alpha$ is considered to be positive. From such energy density we can obtain the DE pressure imposing an EoS for this fluid. Given the $\rho_d$ we can find the pressure to be
\begin{equation}
    P_d = -\rho_d + f(\rho_d) \ ,
\end{equation}
the choice of the DE model will be formalized in the choice of the function $f(\rho)$. There are many options, but we will analyze only two of them:
\begin{itemize}
    \item \underline{Standard $\Lambda$CDM model} \\
    The presence of a cosmological constant in the field equations leads to a effective fluid with $f(\rho_d) = 0$, so that
    \begin{equation}\label{LambdaCDM_EOS}
        P_d = - \rho_d \ .
    \end{equation}
    The change in the star configuration will be in a net smaller pressure compared to the standard GR case, so that the presence of DE will softener the matter distribution.
    \item \underline{$f(\rho) \propto \rho^m$} \\
    The model $f(\rho) \propto \rho^m$ was firstly proposed in \cite{Nojiri_2004} and can be a good model that reproduce different dynamics of a universe filled with DE maintaining an easy analytical expression. Depending on the value of the exponent $m$, we can find different fates for the late time behavior of the universe \cite{_tefan_i__2005,Nojiri_2005}.
    We will choose $m=2$, so that
    \begin{equation}
        f(\rho_d) = \beta \rho_d^2 \ ,
    \end{equation}
    which gives a finite time singularity for $t \rightarrow \infty$ (type III singularity, $\rho$ and $P$ diverge). For positive $\beta$ we have a quintessence model and for negative ones we find phantom DE. The dimension of $\beta$ varies with $m$ and in this case it will be expressed in units of $\beta_0 = (25 \times 10^{14} \text{ g/cm}^3 )^{-1}$.
\end{itemize}
\section{I-Love-Q relations} \label{sec:I-Love-Q}
In this section, we will study the validity of the I-Love-Q relations in the case of a NS with DE contribution. We focused on only two different EoS and realized different combinations of the DE parameters. In particular, we chose the SLy and FPS equations of state. Both of them are tabulated EoS and can be obtained from many-body simulations of the NS interior \cite{Haensel:2004nu}. \\ 
\begin{figure}[H]
    \centering
    \hspace{1.cm}
    % First row with two minipages
    \begin{minipage}[b]{0.4\textwidth}
        \centering
    \includegraphics[width=1.\linewidth]{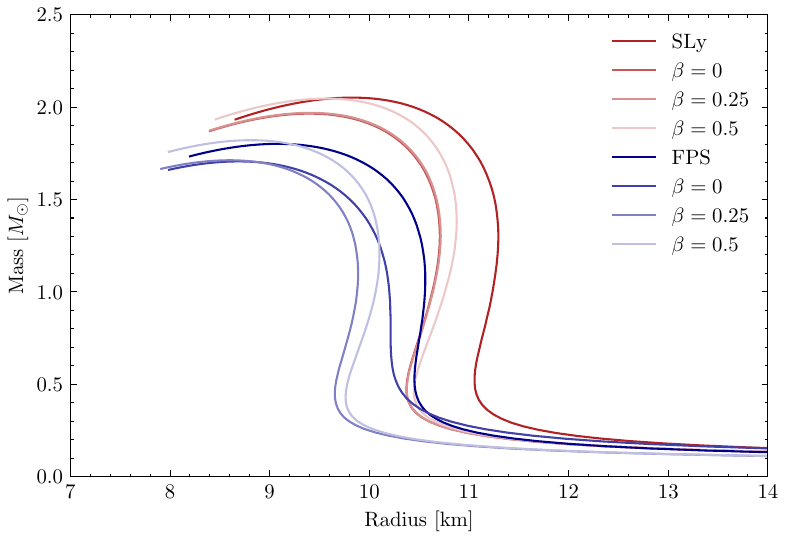}
    \end{minipage}%
    \hfill
    \begin{minipage}[b]{0.4\textwidth}
        \centering
    \includegraphics[width=1.\linewidth]{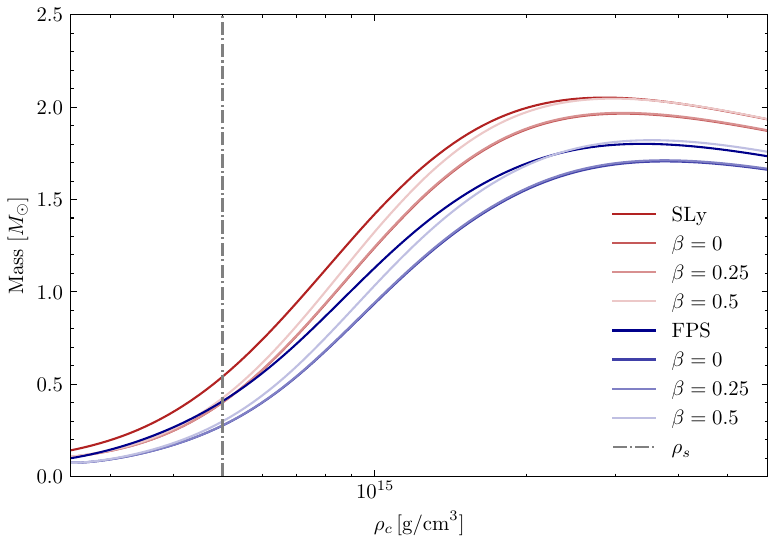}
    \end{minipage}%
    \hspace{1.cm}
    \par
    \hspace{1.cm}
    % First row with two minipages
    \begin{minipage}[t]{0.4\textwidth}
        \centering
    \caption{Mass-Radius plot for the SLy and FPS EoS with and without DE contribution. In all cases we fixed $\alpha = 0.025$ and then we studied for different values of $\beta$ (expressed in units of $\beta_0)$.}
    \label{fig:Mass_radius}
    \end{minipage}%
    \hfill
    \begin{minipage}[t]{0.4\textwidth}
        \centering
    \caption{Plot of the Mass vs central density $\rho_c$ for the SLy and FPS EoS with and without DE contribution. The vertical line represent the value of the DE cutoff $\rho_s =5 \times 10^{14}  \text{g/cm}^3$. 
    In all cases, we fixed $\alpha = 0.025$ and then we studied for different values of $\beta$ (expressed in units of $\beta_0)$.}
    \label{fig:Mass_rho}
    \end{minipage}%
    \hspace{1.cm}
\end{figure}

\begin{figure}[H]
    \centering
    % First row with two minipages
    \begin{minipage}[b]{0.3\textwidth}
        \centering
    \includegraphics[width=1.\linewidth]{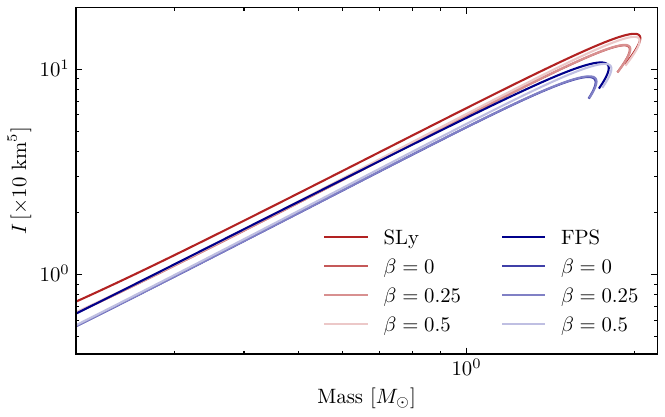}
    \end{minipage}%
    \hfill
    \begin{minipage}[b]{0.3\textwidth}
        \centering
    \includegraphics[width=1.\linewidth]{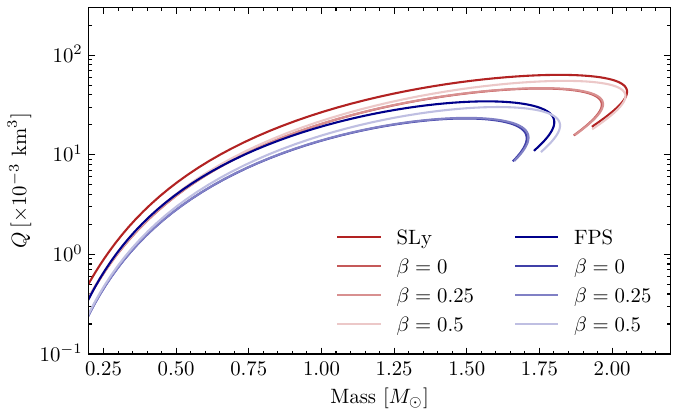}
    \end{minipage}%
    \hfill
    \begin{minipage}[b]{0.3\textwidth}
        \centering
    \includegraphics[width=1.\linewidth]{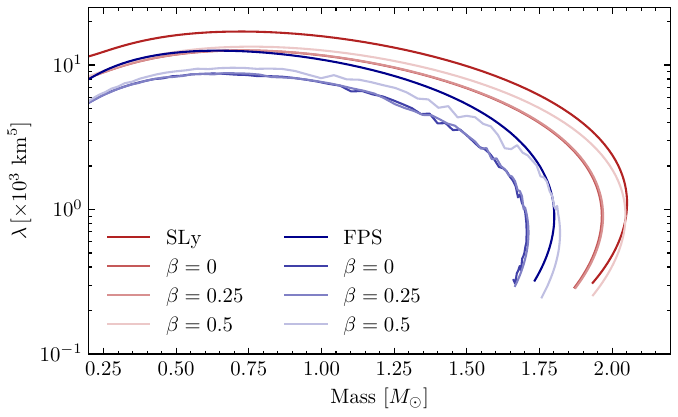}
    \end{minipage}%
    \par
\begin{minipage}[t]{.3\textwidth}
    \caption{Mass vs moment of inertia plot $I$ for the same curves as Figure \ref{fig:Mass_radius}.}
    \label{fig:M_I}
\end{minipage}%
    \hfill
\begin{minipage}[t]{.3\textwidth}  
    \caption{Mass vs quadrupole moment $Q$ plot for the same curves as Figure \ref{fig:Mass_radius}.}
    \label{fig:M_Q}
\end{minipage} %
    \hfill
\begin{minipage}[t]{.3\textwidth}  
    \caption{Mass vs tidal Love number $\lambda$ for the same curves as Figure \ref{fig:Mass_radius}.}
    \label{fig:M_Love}
\end{minipage}   %   
\end{figure}

\begin{figure}[H]
    \centering
    % First row with two minipages
    \begin{minipage}{0.49\textwidth}
        \centering
        % Include your first image
        \includegraphics[width=\textwidth]{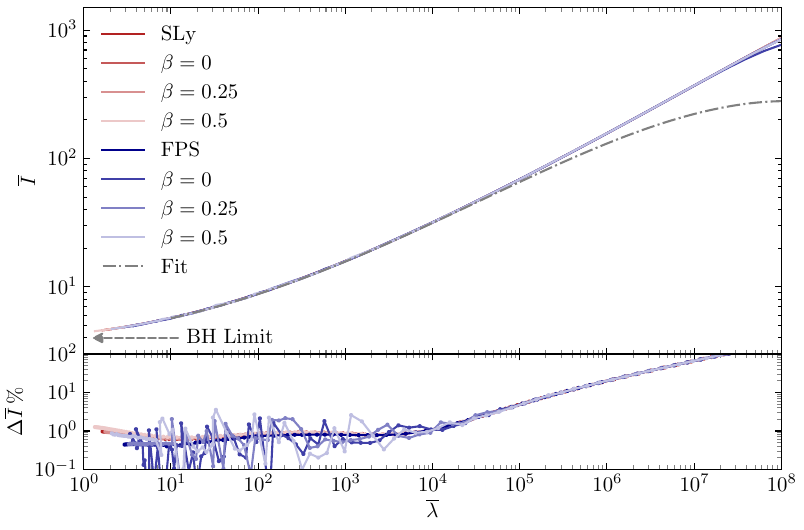}
        \caption{}
        \label{fig:I_Love}
    \end{minipage}%
    \begin{minipage}{0.49\textwidth}
        \centering
        % Include your second image
        \includegraphics[width=\textwidth]{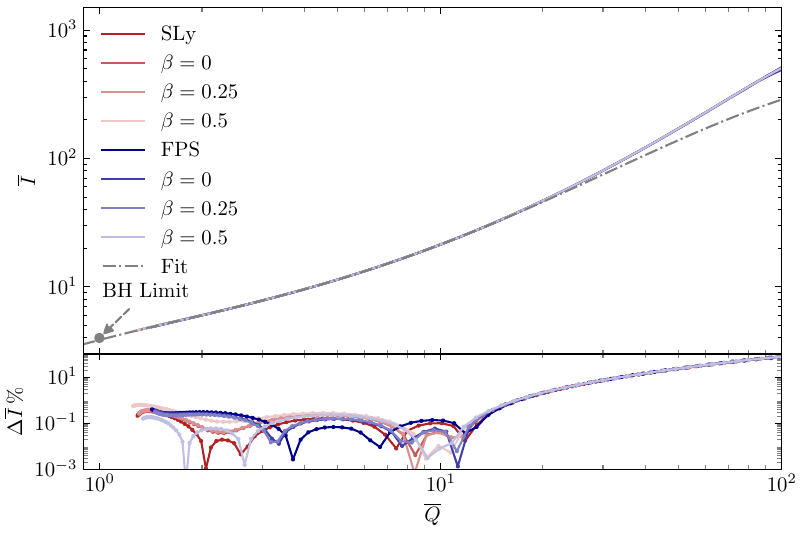}
        \caption{}
        \label{fig:I_Q}
    \end{minipage}
    % Second row with one centered minipage
    
    \vspace{0.5cm} % Adjust vertical space between rows
    \begin{minipage}{0.49\textwidth}
        \centering
        % Include your third image
        \includegraphics[width=\textwidth]{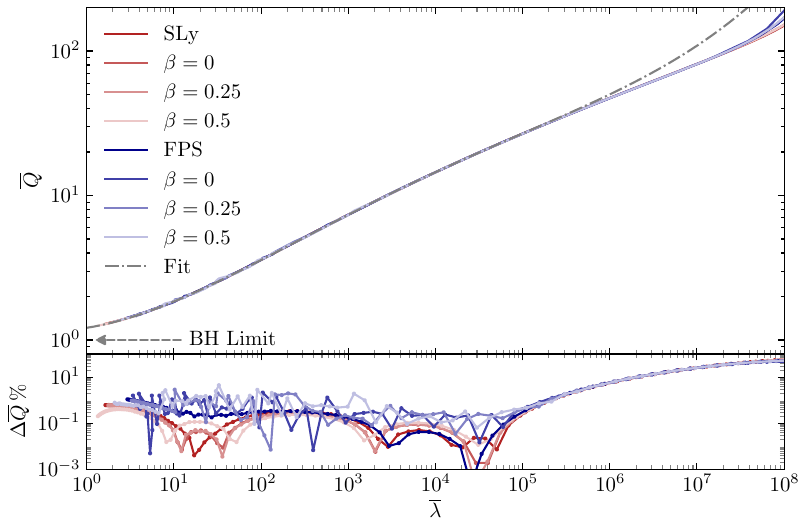}
        \caption{}
        \label{fig:Q_Love}
    \end{minipage}
    \caption*{\centering I-Love-Q plots for the SLy and FPS EoS with and without DE contribution. In all cases we fixed $\alpha = 0.025$ and then we studied for different values of $\beta$ (expressed in units of $\beta_0)$. In all plot we have the fit curve given by Eq. \eqref{fit}. The lower panel of each plot shows the percentage deviation of each dataset from the fit reference curve. In each plot we reported the BH limit, which corresponds to $\bar{I} = 4$, $\bar{Q}=1$ and $\bar{\lambda} = 0$}.
\end{figure}

In Figure \ref{fig:Mass_radius} we can see the mass-radius plot of the different choices of EoS. Firstly, we see the difference between the SLy and the FPS ones, the first gives stars with larger mass and radius, the second one is a softer EoS and gives the same curve for smaller values $M$ and $R$. We then introduced the coupling with DE, which are the curves labeled with different $\beta$s. For illustrative calculation, we decided to fix $\alpha = 0.025$ for all the cases and then study the behavior for different values of $\beta$. The case $\beta = 0$ refers to the standard $\Lambda$CDM model with DE EoS \eqref{LambdaCDM_EOS}. In this case, we see that the introduction of a DE component is softening the EoS, lowering the curve to smaller values of $M$ and $R$, as it could be expected since this coupling results in a lowering of the pressure for a given density. This effect is linear in $\alpha$ \cite{astashenok2023compactstarsdarkenergy} and so we decided only to study one case. The cases for $\beta \neq 0$ seem more interesting since the effect of DE contributes to the diagram differently at different mass ranges, resulting in a change in the shape of the curve. We see, for example for the SLy case, that the deviation starts from around $M \sim M_\odot$, which corresponds to central densities of the order of the cut-off $\rho_s$, as can be seen from the plot in Figure \ref{fig:Mass_rho}. We also notice that in this case the maximum mass increases with $\beta$. This is due to the fact that in this case the DE component is a positive contribution to the pressure. \\
In Figures \ref{fig:M_I}, \ref{fig:M_Q} and \ref{fig:M_Love}, we can see how, respectively,  the moment of inertia $I$, the quadrupole moment $Q$ and the tidal Love number $\lambda$ behave as functions of the mass for the different choices of the EoS. These quantities have been computed using the Eqs. \eqref{I}, \eqref{Q} and \eqref{Love}. In these plots, we see no appreciable changes in the curves' shapes, but there is some shift in them for the different values of $\beta$. The plot of the tidal Love number, Figure \ref{fig:M_Love}, shows a bigger variation for the different cases, and since this parameter is a parameter of the waveform of GWs, this shift can be important for future detections. From the same plot, it seems that $\lambda$ starts to oscillate for values of $\beta \neq 0$. This oscillation starts from the value of mass which is equivalent to $\rho_c >\rho_s$, so where the effects of the DE component are appreciable. We made some tests and it seems these oscillations are not due to the precision of the integration but it is somehow a property of the EoS, maybe due to the fact that the EoS curve has some discontinuities in the first derivative.\\
At this point we can show that these three quantities, $I$, $Q$ and $\lambda$, respect a set of universal relations which are almost independent of the EoS also with the inclusion of a DE component. In order to compare these three quantities we need to adimensionalize them. The adimensional ones, indicated by an over bar, are obtained through a dimensional analysis as \cite{Yagi_2013,Yagi_2013_Science}
\begin{align}
    \bar{I}\equiv& \frac{I}{M^3} \label{barI} \\
    \bar{Q} \equiv& -\frac{Q}{MS^2} \label{barQ}\\
    \bar{\lambda} \equiv& \frac{\lambda}{M^5}\label{barLove} \ .
\end{align}\\
For the analysis of the I-Love-Q relations, we compare the obtained curve with the fit curve 
\begin{equation}\label{fit}
    \ln{y_i} = a_i + b_i \ln{x_i}+ c_i \left(\ln{x_i}\right)^2 +d_i \left(\ln{x_i}\right)^3 + e_i \left(\ln{x_i}\right)^4 \ ,
\end{equation}
where the $i$ refers to the different plots, $\bar{I}-\bar{\lambda}$, $\bar{I}-\bar{Q}$ and $\bar{Q}-\bar{\lambda}$. The values of the coefficients can be found in Table I of \cite{Yagi_2013}. We used these curves to estimate the validity of the relations, in all the plots (Figures \ref{fig:I_Love}, \ref{fig:I_Q} and \ref{fig:Q_Love}) the lower panel shows the deviation of the curves from this reference one. In all cases, the deviation oscillates around $0.1\%$ and $1 \%$ in the range of validity of the fit. This indicates that the I-Love-Q relations are able to describe such objects with reasonable
accuracy.  In each plot, we reported the BH limit of each variable, which is $\bar{I}=4$, $\bar{Q}=1$ and $\bar{\lambda}=0$. One cannot obtain this case by increasing the central density of the object. \\
From both Figure \ref{fig:I_Love} and \ref{fig:Q_Love} we can see that the SLy equation of state provides more stable results when we introduce the DE contribution, since the curves almost stick one to the other. The case of FPS is different and we can see that in the absence of the DE contribution the curve is quite regular, but with such contribution we get some oscillation for small values of $\bar{\lambda}$. These values correspond to large values of mass and so where the coupling with DE becomes relevant, so that it seems that in this case such coupling introduces some instability of the solutions. In the case of Figure \ref{fig:I_Q} we find a good result for all the cases, probably since $\bar{I}$ is a first-order quantity and it is less influenced by modifications of the EoS profile. As already said, in general we obtain reliable results that indicate that the I-Love-Q relations are valid also for NS with DE contributions of this type to the EoS, making even more general and EoS independent such relations. \\

\begin{figure}[H]
    \centering
    % First row with two minipages
    \begin{minipage}[b]{0.3\textwidth}
        \centering
    \includegraphics[width=1.\linewidth]{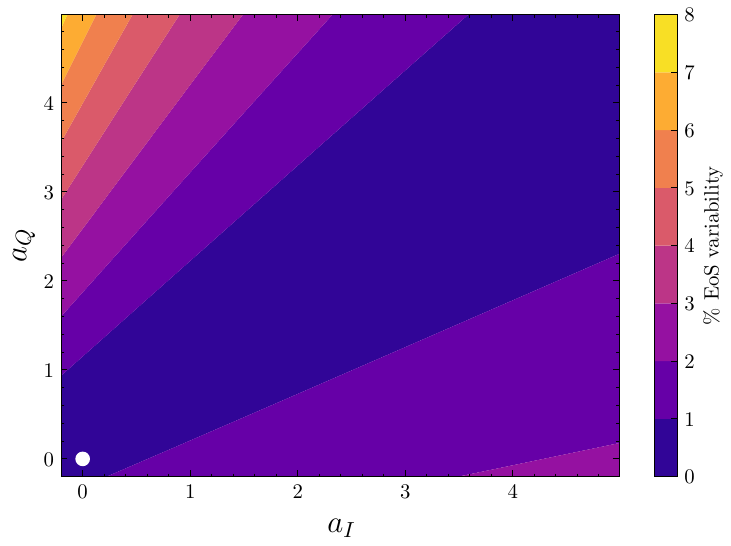}
    \end{minipage}%
    \hfill
    \begin{minipage}[b]{0.3\textwidth}
        \centering
    \includegraphics[width=1.\linewidth]{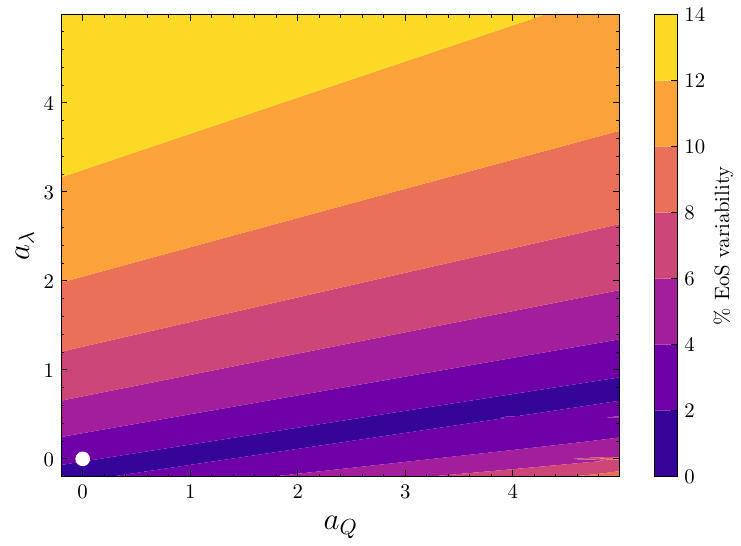}
    \end{minipage}%
    \hfill
    \begin{minipage}[b]{0.3\textwidth}
        \centering
    \includegraphics[width=1.\linewidth]{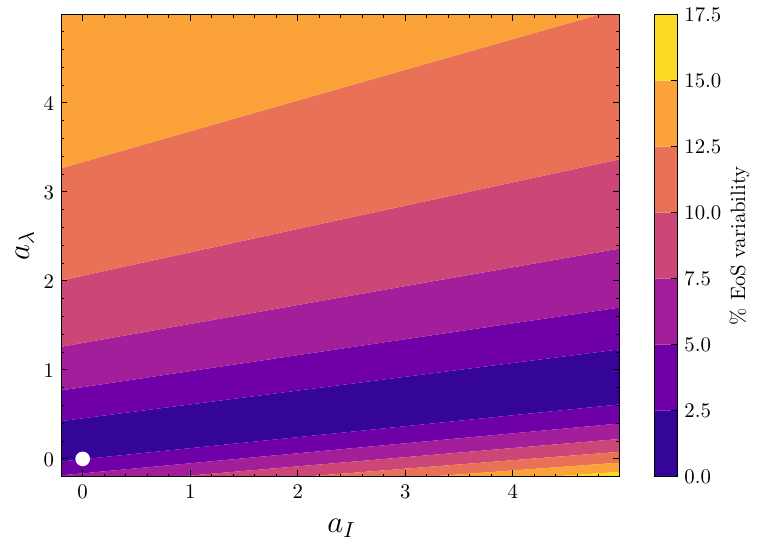}
    \end{minipage}%
    \par

\begin{minipage}[t]{1.\textwidth}
    \caption{EoS variability for different normalization choices in the $I-Q$ (left), $Q-\lambda$ (center) and $I-\lambda$ (right) relations. The white point represent the normalization choice of \cite{Yagi_2013,Yagi_2013_Science}, which is   $(a_I,a_Q,a_\lambda)=(0,0,0)$. The EoS variability has been computed as the maximum percent deviation of the FPS EoS from the SLy one.}
    \label{fig:different_normalizations}
\end{minipage}   %   
\end{figure}

The normalization of the moments \eqref{barI}, \eqref{barQ} and \eqref{barLove} performed in \cite{Yagi_2013,Yagi_2013_Science} is not unique. We still have the freedom to divide each quantity by a power of the compactness as \cite{Yagi_2017,Majumder_2015}
\begin{align}
    \bar{I}^{(a_I)} \equiv \frac{\bar{I}}{\mathcal{C}^{a_I}} \hspace{2cm} \bar{Q}^{(a_Q)} \equiv \frac{\bar{Q}}{\mathcal{C}^{a_Q}}   \hspace{2cm} \bar{\lambda}^{(a_\lambda)} \equiv \frac{\bar{\lambda}}{\mathcal{C}^{a_\lambda}} \ .
\end{align}
The normalization used by \cite{Yagi_2013,Yagi_2013_Science} is then $(a_I,a_Q,a_\lambda)=(0,0,0)$.
As pointed out in \cite{Majumder_2015,Yagi_2017}, for different normalizations, the EoS variability of the universal relations varies. The EoS variability can be estimated by taking the maximum variation of each relation for a set of different EoS. Therefore, there could exist a choice of 
the normalization parameters that minimizes this quantity, leading to more accurate relations. In \cite{Majumder_2015} it has been shown that the best option of these parameters can decrease the variability by a factor of two or more. 
In Figure \ref{fig:different_normalizations} it is reported the EoS variability of each relation, $I-Q$ , $Q-\lambda$ and $I-\lambda$. For each relation, the EoS variability is computed as the maximum percent deviation of the FPS case from the SLy one. The white dot represents the choice $(a_I,a_Q,a_\lambda)=(0,0,0)$ of \cite{Yagi_2013,Yagi_2013_Science}. From this plot we deduce that, also in this case, this choice is not the one that minimizes the EoS variability, and that for the best normalization choice one could reduce the variability by a factor of two, as obtained in \cite{Majumder_2015}.

\section{Conclusions}\label{sec:conclusions}
Many are the theoretical models that provide an explanation for the DE and all of them have been deeply tested in order to reproduce the cosmological observations. Of growing interest is also the attempt to constrain such models from other sources, such as astrophysical ones. The precision of future experiments will provide a good laboratory to test these models. In particular, it is interesting to look at the effect of DE on compact objects. The analysis of these objects can lead to discrepancies or agreements with the standard model that describes them which can help the cosmologist to discriminate between DE models that are accepted from cosmological observations. 
For these reasons, in this paper, we analyzed models on NSs composed of dense matter and DE in the form of a simple fluid. We showed how the effect of such a new component changes the structure of the NS and its properties. We provided for the first time the moment of inertia, the quadrupole moment and the tidal Love number for such objects. This trio of observables will be measured with high precision in future experiments and the presence of DE could be detected from such studies. Then we pointed out how the universal relations still hold for such an interacting fluid, giving a confirmation of the universality of such relations. This relation will be useful to perform redundancy tests and to combine multiple observations. In the future it will be interesting to apply the same analysis to other DE models, using different interactions with  ordinary matter or different DE fluids.  
 Among the possible DE models that can be studied with this formalism there are all the ones that can be described by an effective fluid description, for example entropic DE or $w_0w_a$CDM. Growing interest for this models are arising from the analysis of the recent DESI data. Future data provided by Euclid could in principle confirm the deviations to the standard $\Lambda$CDM model, bringing even more attention on the topic. The possibility to detect effects of such models from astrophysical sources could break degeneracies arising from the only cosmological test of such models. The extension of the analysis to anisotropic stars could be relevant, since the possibility of a breakdown of such relations can occur. In this case we would be also able to describe a larger class of stars, giving the possibility to study DE interactions with more realistic stars or exotic compact objects. Another possibility can be the study of this model in the rapid rotating case in order to see if the DE interaction could alleviate the deviations of the relations.
Also the extension to axions will be important, since they represent good candidates that can solve the dark matter and dark energy problem at the same time. Recent studies have been already performed to study NSs with such component \cite{Astashenok_2020,Astashenok_2020_supermassive} and it will be interesting to check the universal relations of such types of NSs.

\section*{Acknowledgments}
This work is funded by MCIN/AEI/10.13039/501100011033 and FSE+, reference PRE2021-098098 and the program Unidad de Excelencia María de Maeztu
CEX2020-001058-M.

\appendix
\section{Slow-rotating NS}\label{sec:appendix}
In this section, the important equation used to compute the I-Love-Q relations will be introduced, and the numerical procedure followed to integrate such equations will be explained. The analysis performed to integrate the fluid equations has been done mostly following \cite{Yagi_2013}.
\subsection{Zero-order solutions}
At zero-order in rotation, the Einstein equations and the conservation equation for the metric \eqref{HTmetric} give the following system of ODEs
\begin{align}
    \frac{dm}{dr} &= 4 \pi r^2 \rho_0 \label{TOV_Zero_m} \\
    \frac{d\nu}{dr} &= \frac{8 \pi r^3 P_0 + 2m}{r(r-2m)} \label{TOV_Zero_nu}\\
    \frac{dP_0}{dr} &= -\frac{(4 \pi r^3 P_0 + m)(\rho_0+P_0)}{r(r-2m)} \ . \label{TOV_Zero_P} 
\end{align}
These are the equations for the interior of the star. The exterior ones can be obtained fixing $\rho = P = 0$.
This system, along with an EoS for the fluid, can be integrated given the boundary conditions. We will integrate it starting from a given central density $\rho_c$, which will give a corresponding central pressure $P_c$. So that the boundary conditions are given by
\begin{align}
    \rho_0(r_\epsilon) &= \rho_c + \mathcal{O}( r_\epsilon^2) \\  
    P_0(r_\epsilon) &= P_c + \mathcal{O}( r_\epsilon^2) \\ 
    m(r_\epsilon) &= \frac{4}{3} \pi r_\epsilon^3\rho_c + \mathcal{O}( r_\epsilon^5) \\
    \nu(r_\epsilon) &= \nu_c + \mathcal{O}( r_\epsilon^2) \ . 
\end{align}
The condition on $\nu$ is done such that it matches the exterior solution 
\begin{equation}
    \nu^\text{ext}(r) = \ln{\left(1-\frac{2M}{r}\right)} \ ,
\end{equation}
where $M$ is the total mass of the star defined as $m(R)$, where $R$ is the radius of the star. The radius is chosen to be the $r$ such that the pressure drops to $P_0(r)/P_c < 10^{-7}$. In order to match the interior and the exterior solution for $\nu(r)$, we have the freedom of changing the constant $\nu_c$. Since the equation for $\nu'(r)$ \eqref{TOV_Zero_nu} is shift-invariant, we decide to integrate the equation for $\nu_c=1$ and, once $M$ and $R$ have been computed, we shift the value of $\nu_c$ to obtain the right value at $R$.

\subsection{First-order solutions}
\begin{figure}[H]
    \centering
    \includegraphics[width=0.5\linewidth]{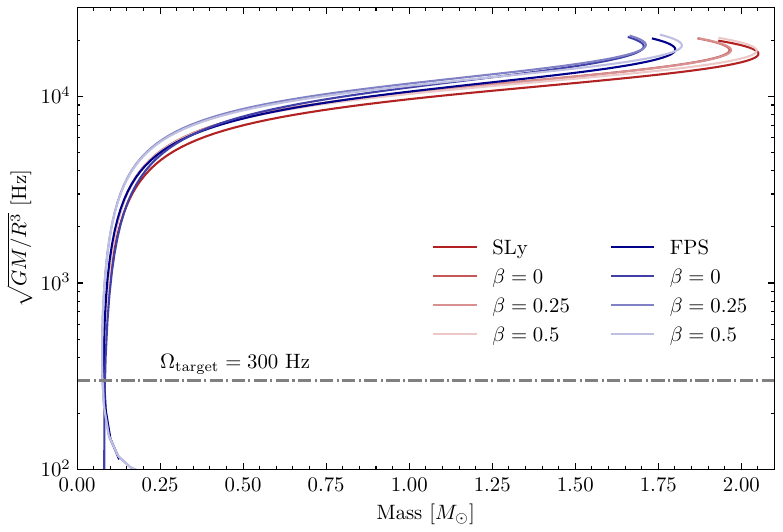}
    \caption{Limiting case of the slow rotating condition for each case of analysis, the chosen angular velocity of the stars respects the condition $\Omega_\text{target} \ll \sqrt{GM/R^3}$ for almost all the central densities.}
    \label{fig:mass_frequency}
\end{figure}

At first order, we only have the contribution from the rotation of the star. This is given by its angular velocity function $\bar{\omega} \equiv \Omega - \omega$, and its evolution is given by the differential equation 
\begin{equation}\label{ODEOmega}
    \frac{d^2\bar{\omega}}{dr^2} + 4 \frac{1-\pi r^2 (\rho_0+P_0)}{r (1-2m/r)} \frac{d\bar{\omega}}{dr} - \frac{16 \pi (\rho_0+P_0)}{ 1-2m/r}\bar{\omega} = 0 .
\end{equation}
Expanding the equation at small $r$ we find the initial conditions to be
\begin{align}
     \bar{\omega}(r_\epsilon) &= \bar{\omega}_c + \frac{8\pi}{5}(\rho_c+P_c)\bar{\omega}_c r_\epsilon^2 +\mathcal{O}( r_\epsilon^3) \\  
    \bar{\omega}'(r_\epsilon) &= \frac{16\pi}{5}(\rho_c+P_c)\bar{\omega}_c r_\epsilon +\mathcal{O}( r_\epsilon^3) \ ,  
\end{align}
where we introduced a single constant $\bar{\omega}_c$. This determines the angular velocity of the star, which may vary from configuration to configuration. The external solution for this function is related to the angular velocity and the spin $S$ of the star as
\begin{equation}
    \bar{\omega}^{\text{ext}}(r) = \Omega - \frac{2 S }{r^3} \ .
\end{equation}
Matching the exterior and interior solutions we find $S = \frac{1}{6}R^4 \bar{\omega}'(R)$ and $\Omega = \bar{\omega}(R)+\frac{2S}{R^3}$. Once we compute spin and angular velocity the moment of inertia of the star $I$ is found to be
\begin{equation}\label{I}
    I= \frac{S}{\Omega} \ . 
\end{equation}
Since the differential equation \eqref{ODEOmega} is scale invariant, we can compute the configuration for stars with the same angular velocity $\Omega_\text{target}$ just integrating the equations and then changing the initial condition as
\begin{equation}
    \bar{\omega}_c \rightarrow \bar{\omega}_c \frac{\Omega_\text{target}}{\Omega} \ .
\end{equation}
The target frequency of rotation has been chosen to be $\Omega_\text{target} = 300 \, \text{Hz}$.
We can see from Figure \ref{fig:mass_frequency} that this value respects the slow rotating condition $\Omega_\text{target} \ll \sqrt{\frac{GM}{R^3}}$ for almost all the central densities.
\\

\subsection{Second-order solutions}
At second order in spin, we studied both the rotating solutions and the tidal one. In the rotating case, the deformation of the star is given just by the fact that the NS is spinning with an angular velocity $\Omega$. In the tidal case, the NS is not spinning and its deformation is generated from an external tidal field, which can be due to a binary companion.

\subsubsection{Rotating case}
The second-order equations for the rotating case have been first introduced by Hartle-Thorne \cite{Hartle:1967he,Hartle:1968si}, a review is given in \cite{Urbanec_2013}. 
Since we are interested in computing the quadrupole moment $Q$ of the star, we will only focus on the perturbations of the metric associated with the $l=2$ angular dependence of the metric, which are $m_2$ , $h_2$ and $v_2$. From the Einstein equations, we find for these functions the differential system
\begin{align}
    \frac{dv_2}{dr} &=-2 \frac{d\nu}{dr} h_2 + \left(\frac{1}{r} + \frac{d\nu}{dr}\right)\left[\frac{1}{6}r^4 j^2 \left(\frac{d\bar{\omega}}{dr}\right)^2 - \frac{1}{3}r^3 \bar{\omega}^2 \frac{d(j^2)}{dr}\right]\label{ODEv2}\\
    \frac{dh_2}{dr} &= -\frac{2v_2}{r(r-2m)d\nu/dr} + \left\{-2\frac{d\nu}{dr} + \frac{r}{2(r-2m)d\nu/dr}\left[ 8\pi(\rho_0+P_0) - \frac{4m}{r}\right]\right\}h_2 \nonumber \\
     & + \frac{1}{6}\left[r\frac{d\nu}{dr} - \frac{1}{2(r-2m)d\nu/dr}\right]r^3 j^2 \left(\frac{d\bar{\omega}}{dr}\right)^2-\frac{1}{3}\left[r\frac{d\nu}{dr}+\frac{1}{2(r-2m)d\nu/dr}\right]r^2\bar{\omega}^2\frac{d(j^2)}{dr} \label{ODEh2}\\
    \frac{m_2}{r-2m} &= -h_2 + \frac{1}{6}r^4 j^2\left(\frac{d\bar{\omega}}{dr}\right)^2 -\frac{1}{3}r^2 \bar{\omega}^2\frac{d(j^2)}{dr} \ ,
\end{align}
where we introduced the function $j(r) = e^{-\nu/2}\sqrt{1-2m/r}$. To integrate this system it is convenient to split the solution into a complementary and a particular part as \cite{Hartle:1967he}
\begin{align}
    h_2 &= h_2^{\text{(P)}} + A h_2^{\text{(C)}} \\
    v_2 &= v_2^{\text{(P)}} + A v_2^{\text{(C)}} \ ,
\end{align}
where $A$ is a constant that will be determined matching the internal and external solutions. The complementary solution is the solution in absence of the rotational part, which it is solution to the system
\begin{align}
    \frac{dv^{\text{(C)}}_2}{dr} &=-2 \frac{d\nu}{dr} h^{\text{(C)}}_2 \\
    \frac{dh^{\text{(C)}}_2}{dr} &= -\frac{2v^{\text{(C)}}_2}{r(r-2m)d\nu/dr} + \left\{-2\frac{d\nu}{dr} + \frac{r}{2(r-2m)d\nu/dr}\left[ 8\pi(\rho_0+P_0) - \frac{4m}{r}\right]\right\}h^{\text{(C)}}_2 \ ,
\end{align}
and the particular solutions are the solutions to the full equations \eqref{ODEh2} and \eqref{ODEv2}. 
The initial conditions for such system are 
\begin{align}
    h^{\text{(C)}}_2 &= B r_\epsilon^2 + \mathcal{O}(r_\epsilon^3)\\
    v^{\text{(C)}}_2 &= -\frac{2\pi}{3}(\rho_c+3P_c) B r_\epsilon^4 + \mathcal{O}(r_\epsilon^5) \\
    h^{\text{(P)}}_2 &= B' r_\epsilon^2 + \mathcal{O}(r_\epsilon^3)\\
    v^{\text{(P)}}_2 &= -\frac{2\pi}{3}(\rho_c+3P_c) B' r_\epsilon^4 -\frac{2\pi}{3}(\rho_c+P_c) j_c^2 r_\epsilon^4 + \mathcal{O}(r_\epsilon^5) \ ,
\end{align}
the constant of integration $B$ has an arbitrary value and the solution is not affected by this choice. For $B'$ we had to perform a shooting method in order to obtain a particular solution which goes to zero at infinity. \\
Outside of the star, the solutions in the asymptotically mass-centered Cartesian coordinates are given by 
\begin{align}
    h_2^{\text{(P)}} + A h_2^{\text{(C)}} &= \frac{S^2}{MR^3}\left(1+\frac{M}{R}\right) + A'\mathcal{Q}_2^2 (\xi) \label{h2secondOrderExterior}\\
    v_2^{\text{(P)}} + A v_2^{\text{(C)}} &= -\frac{S^2}{R^4} + A'\frac{2 M}{\sqrt{R(R-2M)}}\mathcal{Q}_1^2 (\xi) \label{v2secondOrderExterior}\ ,
\end{align}
where $\mathcal{Q}_n^m$ is the associated Legendre function of the second kind and $\xi \equiv R/M-1$. To be more precise, we have
\begin{align}
    \mathcal{Q}_2^2(\xi) &= \frac{3}{2}(\xi^2-1)\log{\left(\frac{\xi+1}{\xi-1}\right)} - \frac{3\xi^2-5\xi}{\xi^2-1} \\
    \mathcal{Q}_1^2(\xi) &= \sqrt{\xi^2-1}\left[\frac{3\xi^2-2}{\xi^2-1}-\frac{3}{2}\xi\log{\left(\frac{\xi+1}{\xi-1}\right)}\right] \ . 
\end{align}
Once the values of the solutions on the radius of the star have been found, we can match with the exterior one defined above (\eqref{h2secondOrderExterior} and \eqref{v2secondOrderExterior}). This gives a system of two equations in two variables ($A$ and $A'$), which can then be found in terms of $h_2(R)$ and $v_2(R)$. With the solution we can then obtain the quadrupole moment $Q$, defined as the coefficient of the $r^3P_2(\cos \theta)$ term in the Newtonian potential \cite{Hartle:1968si}, as 
\begin{equation}\label{Q}
    Q = - \frac{S^2}{M} - \frac{8}{5}A'M^3 \ .
\end{equation}

\subsubsection{Tidal case}
The rotating case was the analysis of isolated NSs which are deformed due to their own rotation. In the tidal case, we focus on non-rotating NSs which are deformed by a tidal field that can be sourced by a binary companion or other sources. The NS will be immersed in a tidal field, which, focusing only on the $l=2$ polar part of the multipole expansion, will be described by $\mathcal{E}^{(\text{tid})}$. The effect of such a perturbation will be a change in the NS quadrupole moment $Q^{(\text{tid})}$. These two quantities are related to the Newtonian potential in
the star’s local asymptotic rest frame (asymptotically mass-centered Cartesian coordinates, ACMC), which is given by 
\begin{equation}\label{ACMCmetric}
    \frac{1-g_{tt}}{2} = - \frac{M}{R} - \frac{Q^{(\text{tid})}}{R^3} P_2(\theta) + \frac{1}{3}\mathcal{E}^{(\text{tid})}R^2 P_2(\theta) \ .
\end{equation}
At linear order in the perturbation $\mathcal{E}^{(\text{rot})}$, the response of the star will be proportional to it, 
\begin{equation}
    Q^{(\text{rot})} = -\lambda \mathcal{E}^{(\text{rot})} \ ,
\end{equation}
where $\lambda$ is the tidal Love number (we stick to the notation of \cite{Yagi_2013}). From it we can derive the tidal apsidal constant
\begin{equation}
    k_2 \equiv \frac{3}{2}\frac{\lambda}{R^5} \ .
\end{equation}
In order to compute the tidal Love number we will need to focus on the non-rotational case, so we take the metric \eqref{HTmetric} and we neglect the rotation contributions ($\bar{\omega}=m_2=k_2=0$). We are then left with only the $h_2$ perturbation function.
Once the solution for such function has been found we compare it to the ACMC expansion of the metric \eqref{ACMCmetric} and extract the Love number. The differential equation defining the perturbation $h_2$ is 
\begin{equation}
    \frac{d^2h_2}{dr^2} + \left\{\frac{2}{r} + \left[\frac{2m}{r} + \frac{4\pi (P_0-\rho_0)}{r-2m}\right]\right\} \frac{dh_2}{dr} - \left\{\frac{6 - 4\pi(5\rho_0+9P_0(\rho_0+P_0)d\rho_0/dP_0}{r(r-2m)} + \left(\frac{d\nu}{dr}\right)^2\right\}h_2=0 \ .
\end{equation}
We can solve this equation with arbitrary initial conditions since the tidal Love number will be scale invariant. Expanding the equation at small $r$ we find
\begin{align}
    h_2 =& a_0 r_\epsilon^2 + \mathcal{O}(r_\epsilon^3) \\
    h_2' =& 2 a_0 r_\epsilon + \mathcal{O}(r_\epsilon^3) \ . 
\end{align}
The tidal Love number will be independent of the integration constant, so that its choice is not important for the following analysis.
Once the solution has been found we can use the general result for NS which will be dependent only on the compactness and the value of the perturbation and its derivative computed on the radius of the star, which is \cite{Hinderer_2008}
\begin{align}
    k_2 =& \frac{8}{5}C^5(1-2C)^2 [2+2C(y-1)-y]\nonumber\\
    &\times\left\{2C[6-3y+3C(5y-8) + 4C^3[13-11y+C(3y-2)+2C^2(1+y)]\right.\nonumber\\
    &\left.\hspace{.5cm}+3(1-2C)^2[2-y+2C(y-1)]\ln{(1-2C)}] \right\}^{-1} \ ,
\end{align}
where $C=M/R$ is the compactness and $y\equiv R h_2'(R)/h_2(R)$.
The polar tidal Love number is then found as
\begin{align}\label{Love}
    \lambda=& \frac{16}{15}R^5C^5(1-2C)^2 [2+2C(y-1)-y]\nonumber\\
    &\times\left\{2C[6-3y+3C(5y-8) + 4C^3[13-11y+C(3y-2)+2C^2(1+y)]\right.\nonumber\\
    &\left.\hspace{.5cm}+3(1-2C)^2[2-y+2C(y-1)]\ln{(1-2C)}] \right\}^{-1} \ .
\end{align}

\phantomsection

\bibliographystyle{apsrev4-2}

\bibliography{Bibliography}

\end{document}